\documentclass[aps,prd,groupedaddress,showpacs,nofootinbib,amssymb,balancelastpage,preprintnumbers,floatfix]{revtex4}

\usepackage[dvipdfmx]{graphicx}
\usepackage{graphicx,bm,color}

\usepackage{amsmath}
\usepackage{amssymb}
\usepackage{amsfonts}
\usepackage{cases}
\usepackage{cancel}
\usepackage[normalem]{ulem}
\usepackage{subfigure}
\usepackage{here}
\usepackage{color}
\usepackage{comment}
\usepackage{hyperref}

\usepackage{tikz}
\usetikzlibrary{matrix}

\allowdisplaybreaks[3]

\begin{document}

\title{ 
Dark QCD perspective inspired by strong CP problem at QCD scale
}

\author{Bin Wang}\thanks{{\tt wangbin22@mails.jlu.edu.cn}}
\affiliation{Center for Theoretical Physics and College of Physics, Jilin University, Changchun, 130012,
	China}

\author{Hiroyuki Ishida}\thanks{{\tt ishidah@pu-toyama.ac.jp}}
\affiliation{Center for Liberal Arts and Sciences, Toyama Prefectural University, Toyama 939-0398, Japan}

\author{Shinya Matsuzaki}\thanks{{\tt synya@jlu.edu.cn}}
\affiliation{Center for Theoretical Physics and College of Physics, Jilin University, Changchun, 130012,
	China}%

\begin{abstract}

We discuss a QCD-scale composite axion model arising from 
dark QCD coupled to QCD. The presently proposed scenario not only 
solves the strong CP problem, but also is compatible with 
the preheating setup for the QCD baryogenesis. 
The composite axion is phenomenologically required to mimic the QCD pion, but 
can generically be flavorful, which could be testable 
via the induced flavor changing processes at experiments. 
Another axionlike particle (ALP) is predicted to achieve the axion relaxation mechanism, which can phenomenologically act as the conventional QCD axion. 
This ALP can be ultralight, having the mass less than 1 eV, to be a dark matter candidate. 
The QCD $\times$ dark QCD symmetry structure constrains dark QCD meson spectra, 
so that the dark $\eta'$-like meson would only be accessible 
at the collider experiments. Still, the Belle II and Electron ion collider experiments can have a high enough 
sensitivity to probe the dark $\eta'$-like meson in the diphoton channel, which 
dominantly arises from the mixing with the QCD $\eta'$ and the pionic composite axion. 
We also briefly address nontrivial cosmological aspects, such as those related to the dark-chiral phase transition, the dark matter production, and an ultraviolet completion related to the ultralight ALP.

\end{abstract}

\maketitle

\section{Introduction}

 QCD is probably the mostly well established sector in the standard model (SM) of particle physics and has so far been extensively explored and well confirmed by collider experiments and lattice QCD, including the clarification of hadron spectra and nonperturbative dynamical features. 
The phenomenology at the QCD scale has thus been well established, which would imply  presence of no extra colored particles.  
 In fact, stringent bounds on the extra light colored particles have been placed by the ALEPH search for gluino and squark pairs tagged with the multijets 
at the LEP experiment~\cite{ALEPH:1997mcm}, which has excluded extra quarkonic and/or gluonic jets with the mass $\gtrsim 10$ GeV. 
However, precise measurements of $\alpha_s$ in lower scales $\lesssim 10$ GeV 
have not been well achieved so far, 
due to the deep infrared complexity of QCD, which is indeed still uncertain~\cite{Deur:2023dzc}. 
Thus, there might be some rooms still left for Beyond the Standard Model (BSM) to be present at the QCD scale of order of hundreds of MeV up to sub GeV, which has conventionally been discarded by event selections at high-energy collider experiments.    
  
When QCD communicates with BSM at the QCD scale via the color interaction, 
the chiral and/or deconfinement phase transition in the thermal history of 
the universe could also be dramatically affected to be substantially altered from 
the conventional scenario; the crossover transition, as has been observed 
in lattice QCD~\cite{Aoki:2009sc,Borsanyi:2011bn,Ding:2015ona,Bazavov:2018mes,Ding:2020rtq,Bazavov:2016uvm,Ding:2017giu}. 
Dark QCD (dQCD) with dark quarks having the QCD colors is generically in the class of many-flavor QCD, so that the dark chiral phase transition would be of first order, which 
is supported from the Pisarski-Wilczek's original work~\cite{Pisarski:1983ms} based on the perturbative renormalization group analysis and also recent functional renormalization group analysis~\cite{Fejos:2024bgl}. 
In that case, the QCD chiral and/or deconfinement phase transition would generically 
be of the type of first order as well, due to the communication with the dark QCD sector. This scenario would also give significant change around the QCD phase transition epoch of the standard Big Bang cosmology. 
Recently, it has been proposed that QCD with such a class of dQCD can provide a central source for the realization of baryogenesis around the QCD phase transition 
epoch, which is what is called like QCD-preheating baryogenesis~\cite{Wang:2022ygk,Wang:2024tda}. 
The induced first-order nature triggers a dynamic, and nonadiabatic fast roll of the quark condensate, monitored by the QCD $\sigma$ meson field, to produce the number densities of particles, in the same way as what the inflaton plays the role, i.e., preheating, in the 
inflationary epoch~\cite{Dolgov:1989us,Traschen:1990sw,Kofman:1994rk,Shtanov:1994ce,Kofman:1997yn} 
(for reviews, see e.g., \cite{Kofman:1997yn,Amin:2014eta,Lozanov:2019jxc}). 
It has also been clarified that this new type of baryogenesis can be 
accommodated consistently with the current astrophysical and cosmological 
constraints, without spoiling the successful QCD hadron physics~\cite{Wang:2024tda}.

This type of QCD $\times $ dQCD can also have a high enough sensitivity to be probed and/or tested by 
 the recent and prospected observations of nano hertz gravitational waves~\cite{NANOGrav:2023gor,NANOGrav:2023hfp,NANOGrav:2023hvm,EPTA:2023sfo,EPTA:2023akd,EPTA:2023fyk,Reardon:2023gzh,Reardon:2023zen,Xu:2023wog}. 
Furthermore, the tail of the gravitational wave spectra produced at the QCD phase transition epoch can still be hunted also at other interferometers designated aiming at the higher frequency spectra, like the Laser Interferometer Space Antenna (LISA)~\cite{LISA:2017pwj,Caprini:2019egz}, 
the Big Bang Observer (BBO)~\cite{Corbin:2005ny,Harry:2006fi}, and Deci-hertz Interferometer Gravitational Wave Observatory (DECIGO)~\cite{Kawamura:2006up,Yagi:2011wg}, etc.

In this paper, we extend such a phenomenologically and cosmologically viable QCD $\times$ dQCD scenario by addressing the strong CP problem in the SM 
and discuss the related phenomenological consequences. 
We model the dQCD sector with $3 + 1$ flavors, where $3$ denotes three Dirac fermions forming the QCD color triplet representation, while $1$ the singlet one. 
This setup is exactly the same as the original literature on the proposal 
of composite axion~\cite{Kim:1984pt,Choi:1985cb} arising as the pseudo Nambu-Goldstone boson of the dark chiral symmetry breaking, 
and also corresponds to the minimal structure of 
the more generic modeling addressed in a context of the asymptotically safe 
theory~\cite{Ishida:2021avk}. 
In that sense, the presently proposed model is interpreted as 
a version having the Peccei-Quinn (PQ) scale at the QCD scale $(\sim 100\,{\rm MeV} - 1\,{\rm GeV})$, while  
the earlier works assume much higher scales (e.g., $\sim 10^{10 \mathchar`- 14}$ GeV).

The composite axion minimizes the QCD $\theta$ and dQCD $\theta$ vacua to be free from both QCD and dQCD $\theta$ parameters 
by a hybrid relaxation mechanism together with another axionlike particle (ALP) in a way similar to the literature~\cite{Draper:2018tmh,Draper:2016fsr,Agrawal:2017ksf,Gaillard:2018xgk,Gherghetta:2020ofz,Ishida:2021avk}.  
We find that the QCD-scale composite axion is phenomenologically required to mimic the QCD pion (with the same mass $\sim 140$ MeV and the same lifetime), but can generically be flavorful due to the intrinsically minimal flavor violation as has been explicitly addressed in~\cite{Cui:2021dkr}. 
This pionic axion could be testable via the induced flavor changing processes at experiments, as has been discussed in~\cite{Ishida:2020oxl,Cui:2021dkr,Bhattacharya:2021shk,Bauer:2021mvw,Bandyopadhyay:2021wbb,Yue:2022ash,Zhevlakov:2022vio}.

One of another type of ALPs, to be denoted as $P$, is predicted to act as the conventional QCD axion where 
the mass and the PQ breaking scale are correlated via the QCD scale (the QCD topological susceptibility). 
This ALP can be ultralight, having the mass less than 1 eV, to be a dark matter candidate.

Since the QCD $\times$ dark QCD color symmetry structure constrains dark QCD meson spectra, 
the dark $\eta'$-like mesons (dubbed $\eta_d$) would only be accessible 
at the collider experiments. 
We find that the Belle II~\cite{Dolan:2017osp} and the Electron-Ion Collider (EIC)~\cite{Balkin:2023gya} can have a high enough 
sensitivity to probe the dark $\eta'$-like meson in the diphoton channel, which 
dominantly arises from the mixing with the QCD $\eta'$ and the pionic composite axion. 
Nontrivial cosmological aspects related to the dark-chiral phase transition, including the gravitational wave production, the dark matter production, and an ultraviolet completion related to the ultralight ALP ($P$), are also briefly addressed.

\section{The composite axion model from QCD $\times$ dQCD with $3 + 1$ flavors at the QCD scale}

In this section we start with modeling of the dark QCD sector with 
the QCD-color triplet Dirac-fermions $Q_{L,R}$ and the singlet one $\chi_{L,R}$. 
For the vacuum to be free from any CP phase, we also assume 
the presence of a complex scalar $\phi$, which is polar decomposed into the real scalar $S$ and pseudoscalar $P$ fields 
as $\phi = \frac{S}{\sqrt{2}} e^{i \frac{P}{v_S}}$ with 
the vacuum expectation value (VEV) $v_S$. 
This global $U(1)$ symmetry acts as a part of the PQ symmetry. 
The origin of the spontaneous 
breaking of this $U(1)$ symmetry is briefly to be addressed in the section of Summary and 
discussions. 
The charge assignment regarding the dQCD sector together with 
the complex scalar $\phi$ is given in Table~\ref{tab:charge}.

\begin{table}[t] 
\begin{tabular}{|c|ccc|c|} 
\hline 
& $SU(3)_c$ & $SU(2)_W$ & $U(1)_Y$ & $SU(3)_d$  \\  
\hline \hline 
$Q_{L,R}$  & 3 & 1 & 0 & 3 \\  
$\chi_{L,R}$ & 1 & 1 & 0 & 3 \\ 
\hline 
$S $ & 1 & 1 & 0 & 1 \\ 
\hline 
$P $ & 1 & 1 & 0 & 1 \\ 
\hline 
\end{tabular}
\caption{
The charge assignment table for the dQCD fermions $Q$ and $\chi$, 
and the complex scalar formed from $S$ and $P$ as $\phi = \frac{S}{\sqrt{2}} e^{i \frac{P}{v_S}}$. The subscripts $c, w, Y$, and $d$ stand for the QCD color, the $SU(2)$ weak charge, the hypercharge, and the dQCD color, respectively.   
}
\label{tab:charge}
\end{table}

The strong coupling scale of dQCD, $\Lambda_d$ (called the intrinsic scale), is assumed to be slightly higher than, but still close to the intrinsic scale of QCD $\Lambda_c$: $\Lambda_d \gtrsim \Lambda_c \sim 4 \pi f_\pi (\sim 1\,{\rm GeV})$.  
The dQCD sector posses the chiral $U(4)_L \times U(4)_R$ symmetry acting on $F=(Q, \chi)^T$. This chiral symmetry is approximately present, which is explicitly broken by the QCD singlet fermion $\chi$ coupled to the complex scalar $\phi$, 
    \begin{align}
       - {\cal L}_{g_\chi} =   g_\chi (\phi \bar{\chi}_L  \chi_R + {\rm h.c.})    
        \,, \label{L:S}
    \end{align}
with the Yukawa coupling $g_\chi$ being assumed to be small.  
The VEV of the radial part of the complex scalar $S$ provides the current   
    masses for $\chi$, $m_\chi = g_\chi  \frac{v_S}{\sqrt{2}}$, which are assumed to be $\ll \Lambda_d$ irrespective to $v_S$.

 Including the Yukawa terms in Eq.(\ref{L:S}), 
    the anomalous conservation laws for the QCD-singlet axial currents $J_{\mu5}^0 = \bar{F}  \gamma_\mu \gamma_5 T^0 F$ and 
    $J_{\mu5}^{15} = \bar{F} \gamma_\mu \gamma_5 T^{15} F$ take the form: 
\begin{align} 
     \partial^\mu J_{\mu5}^0 
     &= \frac{2}{\sqrt{8}} i  m_\chi \bar{\chi} \gamma_5 \chi 
     - \frac{g_c^2}{32\pi^2} \frac{3}{\sqrt{8}} G\cdot \tilde{G} 
     - \frac{g_d^2}{32\pi^2} \frac{4}{\sqrt{8}} G_d \cdot \tilde{G}_d
     \,, 
     \notag\\ 
     \partial^\mu J_{\mu5}^{15} 
     &= \frac{3}{2\sqrt{6}} i m_\chi \bar{\chi} \gamma_5 \chi   
     + \frac{g_c^2}{32\pi^2} \frac{3}{2 \sqrt{6}} G\cdot \tilde{G} 
\,, \label{ano:d}
\end{align}
where $G_{(d)}$ denotes the (d)QCD gluon field strength with the gauge coupling $g_{c(d)}$, 
and $G_{(d)}\cdot \tilde{G}_{(d)}$ stands for $\frac{\epsilon_{\mu\nu\rho\sigma}}{2} G^{\mu\nu}_{(d)} G^{\rho\sigma}_{(d)}$. 
In Eq.(\ref{ano:d}), 
$T^0 = \frac{1}{\sqrt{2\cdot 4}}\cdot 1_{4 \times 4}$ and $T^{15} = \frac{1}{\sqrt{2 \cdot 12}}\cdot {\rm diag}(-1,-1,-1,3)$. 
Other currents are completely conserved at the classical level.  
    At the low-energy $\lesssim \Lambda_d$, 
    the (approximate) 
    dark chiral $U(4)_L \times U(4)_R$ symmetry is spontaneously broken down to the vectorial subgroup $U(4)_{V=L+R}$ as 
    $U(4)_L \times U(4)_R \to U(4)_V$ with the sixteen Nambu-Goldstone bosons, called the dark pions. 
    The spontaneous breaking is triggered by the nonperturbatively-generated dQCD chiral-condensate $\langle \bar{F}F \rangle$. 
    In the chiral broken phase, the dark fermion-flavor bilinear $\bar{F}F$ can approximately be parameterized as 
    \begin{align}
        \bar{F}_L F_R \approx \langle \bar{F}F \rangle \cdot U_d 
        \,, 
        \qquad 
        U_d = e^{2 i \frac{\pi_d^a}{f_{\pi_d}}T^a} 
        \,, \qquad \pi_d = \sum_{a=0}^{15} \pi_d^a \cdot T^a 
        \,, \label{Ud}
     \end{align}
    where $f_{\pi_d}$ denotes the dark pion decay constant and $T^a$ are generators of $U(4)$ normalized as ${\rm tr}[T^a T^b] = \frac{1}{2}\delta^{ab}$ for $a,b=0, \cdots , 15$. 
    The chiral field $U_d$ transforms in the same way as the dark quark bilinear as $U_d \to g_L ^\dag\cdot U_d \cdot g_R$ 
    with $g_{L,R}$ being the transformation matrices associated with $U(4)_{L/R}$ group.  
    Among six-teen Nambu-Goldstone bosons, the $U(4)$ singlet dark pion $\pi_d^0 
\equiv \eta_d$ gets heavy by feeding the dark $U(1)_A$ anomaly as in the first line of Eq.(\ref{ano:d}), just like $\eta'$ in QCD, hence is to be decoupled at the low-energy. 
In addition, the $SU(3)$ singlet, but $SU(4)$ adjoint dark pion $\pi_d^{15} \equiv a$ gets the mass from the QCD-induced anomaly, via the second line of Eq.(\ref{ano:d}).

The decay constant of the dQCD anomaly-free $a$, denoted as $f_a$, 
is defined as 
\begin{align}
    \langle 0 | \frac{g_c^2}{32 \pi^2} G \cdot \tilde{G} | a(p=0) \rangle = - m_a^2 f_a 
    \,, 
\end{align}
so that with the lowest pole dominance, 
we have the QCD topological susceptibility, 
\begin{align} 
\chi_{\rm top}^{\rm QCD} = \int d^4 x \langle 0 | \left( \frac{g_c^2}{32 \pi^2} G(x) \cdot \tilde{G}(x) \right) \left( \frac{g_c^2}{32 \pi^2} G(0) \cdot \tilde{G}(0) \right)| 0 \rangle \sim \frac{f_a^2 m_a^4}{m_a^2} = f_a^2 m_a^2
\,. \label{chitop}
\end{align} 
Equivalently, we have the QCD gluon-gluon-$a$ interaction of the form  
\begin{align} 
{\cal L}_{aGG} = \frac{g_c^2}{32\pi^2} \frac{a}{f_a} G \cdot \tilde{G}
\, . 
\label{agg}
\end{align} 
From Eq.(\ref{ano:d}) and the definition of the dark pion decay constant $f_{\pi_d}$, 
$\langle 0 | J_\mu^a(x) | \pi_d^b(p) \rangle = - i p_\mu f_{\pi_d} e^{- i px}\cdot \delta^{ab}$, 
we then find  
\begin{align}
    \frac{1}{f_a} = \frac{3}{2 \sqrt{6}} \frac{1}{f_{\pi_d}} 
\,. \label{fa}
\end{align}

\section{Hybrid ALP relaxation} 

Through the QCD and dQCD axial anomalies with the mass terms in Eq.(\ref{L:S}) 
as well as the ordinary QCD quark mass terms, 
in the chiral broken (and confinement) phases for both QCD and dQCD, 
$a$, $P$, $\eta_d$, and the QCD $\eta'$ (in the three-flavor singlet representation) can be mixed in the following induced 
potential terms:  
    \begin{align}
        V_{\eta',\eta_d, a,  P} \Bigg|_{\rm mass} 
        &= 
        2  \left[ \sum_{q=u,d,s} m_q  \langle \bar{q} q \rangle  \cos\left( \frac{2\eta'}{\sqrt{6} f_{\pi}}   \right) +m_\chi  \langle \bar{F}F \rangle \cos\left( \frac{\eta_d}{\sqrt{2} f_{\pi_d}}  + \frac{3 a}{\sqrt{6} f_{\pi_d}} +\frac{P}{v_S} \right) \right]  
        \notag\\
       V_{\eta', \eta_d, a,  P } \Bigg|_{\rm anomaly}^{\rm flavor-universal} 
         & =\chi_{\rm top}^{\rm QCD} \cdot \cos
         \left(\Bar{\theta}_{\rm QCD}+\frac{3 \eta'}{ \sqrt{6} f_{\pi}}+ \frac{3\eta_d} {2\sqrt{2} f_{\pi_d}}-\frac{3a} {2\sqrt{6} f_{\pi_d}}\right)  
         \notag\\ 
         & + \chi_{\rm top}^{\rm dQCD} \cdot \cos \left(\Bar{\theta}_{\rm dQCD}+
         \frac{\sqrt{2} \eta_d}{f_{\pi_d}} \right)  
         \notag\\
         V_{\eta', \eta_d,} \Bigg|_{\rm anomaly}^{\rm flavor-non-universal} &=\frac{1}{2}m_{\eta'}^2 \eta'^2+ \frac{1}{2} m_{\eta_d}^2 \eta_d^2
         \,. \label{V}
    \end{align}
Here are several remarks to note in deriving those potential terms: $\chi_{\rm top}^{\rm dQCD}$ is defined as in the same way as the QCD case, given in Eq.(\ref{chitop}); since the explicit breaking mass $m_\chi$ is assumed to be small enough, the dQCD chiral condensate has been evaluated to be the $SU(4)_V$ symmetric form, i.e., $\langle \bar{\chi}\chi \rangle = \langle \bar{F}F \rangle$ (per dQCD flavor); the dQCD $\theta$ parameter $\theta_{\rm dQCD}$ has been shifted, by absorbing the CP phase of the Yukawa coupling $g_\chi$, to $\bar{\theta}_{\rm dQCD}$, 
as well as the QCD $\bar{\theta}$ which includes the QCD quark CP phase; 
only the flavor-universal part of the QCD axial anomaly, which corresponds to the second line of Eq.(\ref{V}) coupled to $\chi_{\rm top}^{\rm QCD}$, can mix 
the ALPs ($\eta', a, P, \eta_d$) by reflecting the flavor-singlet nature of 
the (d)QCD vacuum~\cite{Baluni:1978rf,Kim:1986ax,Kim:2008hd} (for the recent literature on the flavor-universality of $\chi_{\rm top}$ in QCD, see also~\cite{Kawaguchi:2020kdl,Kawaguchi:2020qvg,Cui:2021bqf,Cui:2022vsr,Huang:2024nbd}); the flavor-non-universal part of the QCD and dQCD axial anomalies respectively 
generate the $\eta'$ and $\eta_d$ masses, as in the last line of Eq.(\ref{V}). 
Other dark pion potential terms, which are anomaly free,  
trivially becomes stationary at $\pi=0$, while 
terms including the current target ALPs ($a, \eta_d, \eta'$ and $P$) are nontrivially relaxed as seen right below.

We rotate the massless dQCD $Q$ quarks by the axial transformation as $Q\to e^{- i \alpha_Q \gamma_5/2} \cdot Q$ with $\alpha_Q=-\bar\theta_{\rm dQCD}$, so that $\Bar\theta_{\rm dQCD}$ is eliminated to be absorbed into the redefinition of $\Bar\theta_{\rm dQCD}$ as 
\begin{align}
    \Bar\theta_{\rm QCD} \frac{g_c^2}{32\pi^2} G\cdot \tilde{G}
    +\Bar{\theta}_{\rm dQCD}\frac{g_d^2}{32\pi^2}  {G_d}\cdot \tilde{G_d}
        \quad \to \quad \Bar\theta_{\rm eff} \frac{g_c^2}{32\pi^2}  G\cdot \tilde{G}
        \,, 
\end{align}
where $\bar\theta_{\rm  eff}=\bar\theta_{\rm QCD}-\bar\theta_{\rm dQCD}$. 
Then the ALP potential in Eq.(\ref{V}) is minimized at 
the vacuum, 
    \begin{align}
    \eta'&=\eta_d=0
    \,,
    \notag\\
    \frac{a}{f_a}&= \bar\theta_{\rm eff}
    \,,
    \notag\\
    \frac{P}{v_S}&= - 2 \bar\theta_{\rm eff}
    \,. 
    \label{vac}
    \end{align}
Thus the hybrid relaxation, which is essentially played by $a$ and $P$, 
works to make the vacuum of dQCD and QCD CP invariant.

The $\eta_d$ mass is supposed to be on the same order as the QCD $\eta'$ mass of $\sim 1$ GeV, but is still slightly higher. 
The $a$- and $P$-ALPs are then mixed more strongly, but still weakly (with 
the size of mixing angle $\sim f_a/v_S \ll 1$) to give 
the mass eigenvalues: 
\begin{align}
    m_a^2 &\simeq \left( 210 \,{\rm MeV} \times 
    \left( \frac{\langle - \bar{F}F \rangle}{\left( 600\,{\rm MeV} \right)^3} \right)^{1/2} 
    \cdot \left( \frac{200\,{\rm MeV}}{f_a} \right)\cdot 
    \left( \frac{m_\chi}{1\,{\rm MeV}} \right)^{1/2} \right)^2 
    \notag\\ 
    & +\left(30\,{\rm MeV} \times \left( \frac{|\chi_{\rm top}^{\rm QCD}|^{1/4}}{76\,{\rm MeV}} \right)^4
    \cdot  \left( \frac{200\,{\rm MeV}}{f_a}  \right)  \right)^2
    \,, 
    \notag\\
    m_P^2 &\simeq \left(0.1\,{\rm eV} \times \left( \frac{|\chi_{\rm top}^{\rm QCD}|}{(76\,{\rm MeV})^4} \right)^{\frac{1}{2}}
    \cdot  \left( \frac{\langle - \bar{F}F \rangle}{\left( 600\,{\rm MeV} \right)^3} \right)^{1/2} \cdot \left( \frac{m_\chi}{1\,{\rm MeV}} \right)^{1/2} \right)^2
    \notag\\
    & \times\left (\left( \frac{200\,{\rm MeV}}{f_a} \right)\cdot\left( \frac{2.8\times{10^5}\,{\rm TeV}}{v_S} \right)\cdot \left( \frac{210\,{\rm MeV}}{m_a} \right) \right)^2
    \,, \label{ma:ref}
\end{align}
 where we have quoted \cite{Borsanyi:2016ksw} for $\chi_{\rm top}^{\rm QCD}$. Thus  the $a$-ALP gets mass mainly from the current $\chi$ mass, i.e., $v_S$, rather than the QCD axial anomaly.

\section{Constraints on ALPs and predictions}

As has been noted in~\cite{Wang:2024tda}, 
the class of the presently employed dQCD does not spoil the currently observed QCD hadron physics, due to the high protection by the double color symmetries: 
QCD colored-dQCD mesons will form exotic QCD hadrons by pairing two, e.g., $(\bar{Q}\chi) \cdot (\bar{\chi} Q)$ and $(\bar{Q} \lambda^a Q) \cdot (\bar{Q} \lambda^a) Q$ (where $\lambda^a$ denote the Gell-Mann matrices), which will be at least as heavy as the QCD scale $\sim 1\, {\rm GeV}$ due to the QCD interaction. Those states will be identified as multijets with the energy around the QCD scale at collider experiments, which is, however, highly challenging to probe at the current detection status as has been noted in the Introduction. 
Still, however, the ALPs ($a, P, \eta_d$) as the QCD color singlets can have a sensitivity to be constrained via the coupling to diphoton, which arises from the $a - \gamma- \gamma$ and 
$\eta' - \gamma - \gamma$, as will be seen more clearly below. 
In this section we discuss the current experimental and astrophysical bounds 
on those ALPs.

\subsection{The ALP: $a$}

The $a$ couplings to the SM quarks and photon are present due to 
the coupling to the QCD $G \cdot G$, which is to be generated via 
the axial rotation as  
\begin{align}
    \frac{a}{f_a} 
    \frac{g_c^2}{32 \pi^2} G \cdot \tilde{G}  \qquad \to \qquad 
  \sum_{q=u,d,\cdots } 
    m_q \bar{q} e^{2i \frac{a}{f_a}\gamma_5}   q - \frac{a}{f_a}\frac{3 \alpha_{\rm em}}{4} \sum_{q=u, d,... } [Q_{\rm em}^q]^2 F \cdot \tilde{F}  
    \,, \label{ALP-couplings}
\end{align}
where $F_{\mu\nu}$ is the photon field strength; 
$\alpha_{\rm em}$ is the fine-structure constant of electromagnetic coupling defined as 
$\alpha_{\rm em} = e^2/(4\pi)$ with the electromagnetic coupling $e$; 
$Q^{\rm em}_{q_i}$ denotes the electric charge (in unit of e) for $i-$ quark.

The $a$-ALP couplings can be flavorful by 
taking into account the Cabibbo-Kobayashi-Maskawa (CKM) mixing in the SM quark mass term, 
\begin{align}
 -   {\cal L}_q^{\rm mass} \to \sum_{i,j,} \bar{q}_i m^{ij}_q  q_j + {\rm h.c.} 
 \,. \label{q-mass-term}
\end{align}
This is called the intrinsic flavorful ALP, or``minimal flavor violatoin (MFV)" for ALP~\cite{Cui:2021dkr}. 
Recall the quark-mass diagonalization in the SM, which is worked out 
by bi-unitary (left and right) transformations of the mass matrices. 
The degree of freedom of the left-handed rotation is completely fixed by 
the CKM,   
while the right-handed one is unfixed within the SM, to be left as 
hidden, redundant and unphysical parameters. 
When the ALP is present and has the chiral couplings to SM fermions, as in Eq.~(\ref{ALP-couplings}), 
the redundant right-handed rotation actually becomes physical. 
This is only the flavor-violating source for the ALP within the ``MFV" framework, 
and breaks the parity, allowing the ALP to couple to SM quarks,  
not only in an axial (pseudoscalar) form, but also in a vector (scalar) form 
as discussed in the literature~\cite{Cui:2021dkr}.

We shall work on the basis where 
the left-handed up-sector quark fields are unchanged, which is possible due to the $SU(2)$ gauge invariance in the SM Yukawa sector, so that the left-handed down-sector quark fields 
are rotated by the CKM matrix as $d^i_L \to V_{\rm CKM}^{ij} \cdot d^j_L$ ($j=1,2,3$ 
corresponding to $d,s,b$). 
The rotation matrix of right-handed quark fields, $U_R^{u}$ and $U_R^{d}$, can be defined in a way similar to the CKM  matrix. 
Those rotation matrices are thus parametrized as 
\begin{equation}
    U_R^{u,d}=\left(\begin{array}{ccc}
c_{12} c_{13} & s_{12} c_{13} & s_{13}  \\
-s_{12} c_{23}-c_{12} s_{13} s_{23}  & c_{12} c_{23}-s_{12} s_{13} s_{23}  & c_{13} s_{23} \\
s_{12} s_{23}-c_{12} s_{13} c_{23}  & -c_{12} s_{23}-s_{12} s_{13} c_{23}  & c_{13} c_{23}
\end{array}\right)^{u,d}
\,, \label{UR}
\end{equation}
where $c_{ij} = \cos \theta_{ij}$ and $s_{ij} = \sin \theta_{ij}$. 
By performing these rotations, 
the $a$-ALP couplings to SM quark mass-dependent terms in Eq.~(\ref{ALP-couplings}) 
together with the SM quark mass terms in Eq.(\ref{q-mass-term})   
are thus cast into the form in the mass eigenstate basis:
\begin{align}
 &  \sum_{u_i = \{ u,c,t\}} \bar{u}_{i} m_{u_i}  u_i 
+ 
\sum_{d_i = \{d,s,b\}} \bar{d}_i m_{d i} d_i  
\notag\\ 
& + 
      i  \sum_{u_i} \frac{a}{f_a}\bar{u}_i \left( (g_V^u)_{ij}+(g_A^u)_{ij} \gamma_5 \right) u_j   
    + i  \sum_{d_i} \frac{a}{f_a}\bar{d}_i \left( (g_V^d)_{ij}+(g_A^d)_{ij} \gamma_5 \right) d_j 
\,,    \label{mass}
\end{align}
where $m_{u_i}$ and $m_{d_i}$ are ${\rm diag}(m_u, m_c, m_t)$ and ${\rm diag}(m_d, m_s, m_b)$, respectively, and 
the vector and axial couplings are given as 
\begin{align}
    g_V^{u} &=\frac{C_{u i} \cdot U_R^u - {U_R^u}^{\dagger} \cdot C_{u i}}{2}, \,, \notag\\ 
    g_A^{u} &=\frac{ C_{u i} \cdot U_R^u + {U_R^u}^{\dagger} \cdot C_{u_i}}{2} 
    \,, \notag\\
    g_V^{d} &=\frac{V_{\rm CKM}^\dag \cdot C_{d i} \cdot U_R^d - {U_R^d}^{\dagger} \cdot C_{d i}\cdot V_{\rm CKM}}{2}, \,, \notag\\ 
    g_A^{d}&=\frac{V_{\rm CKM}^\dag \cdot C_{d i} \cdot U_R^d + {U_R^d}^{\dagger} \cdot C_{d_i} \cdot V_{\rm CKM}}{2} 
    \,, \notag\\  
    C_{u_i} &= 2 m_{u_i} \,,  \qquad C_{d_i} = 2 m_{d_i} 
     \,. \label{gVA}
\end{align}
Then the $a$-ALP photon coupling in Eq.(\ref{ALP-couplings}) 
gets modified as well.  
When the $a$-ALP and photons are onshell, 
including the quark-one loop corrections, 
the coupling reads~\cite{Bauer:2020jbp} 
\begin{align} 
{\cal L}_{a \gamma \gamma} 
&= 
\frac{g_{a\gamma\gamma}}{4} a \cdot F_{\mu \nu} \Tilde{F}^{\mu \nu}, 
\label{a-g-g}
\end{align} 
where 
\begin{align} 
   g_{a\gamma\gamma} &\equiv 
   \frac{\alpha }{\pi f_a} C_{\gamma \gamma}^{\rm eff}  \,, \notag\\ 
   C_{\gamma \gamma}^{\rm eff} & \equiv  
   3 \left[ \sum_{u_i}   (Q_{u_i}^{\rm em})^2 \frac{(g_A^u)_{i i}}{m_{u_i}} B_1 \left(\frac{4m_{u_i}^2}{m_a^2} \right) + 
  \sum_{d_i}   (Q_{d_i}^{\rm em})^2 \frac{(g_A^d)_{i i}}{m_{d_i}} B_1 \left(\frac{4m_{d_i}^2}{m_a^2} \right)
   \right] 
   \,. 
\label{Cgammagamma}
\end{align} 
The loop function $B_1(x)$ is given as 
\begin{equation}
    B_1(x)=1 -xf^2(x), \quad f(x)= 
    \begin{cases} 
    \displaystyle \sin^{-1}\frac{1}{\sqrt{x}} \hspace{24.3mm} {\rm for} \qquad x\geq 1\\ 
    \displaystyle \frac{\pi}{2}+\frac{i}{2} \ln\frac{1+\sqrt{1-x}}{1-\sqrt{1-x}} \hspace{7mm} {\rm for} \qquad x < 1
    \end{cases}
\,. 
\end{equation}

Around sub hundreds of MeV up to sub GeV mass range for the ALP, 
the existing experimental limits on the ALP-photon coupling $g_{a\gamma\gamma}$ in Eq.(\ref{Cgammagamma}) include   
$e^{+} e^{-} \to 3 \gamma$ at Belle II~\cite{Belle-II:2020jti} and BESIII~\cite{BESIII:2022rzz}, 
a photon-beam experiment~\cite{Aloni:2019ruo}; heavy-ion collisions~\cite{CMS:2018erd};  
electron beam dump experiments~\cite{Dolan:2017osp};  
SN 1987A~\cite{Jaeckel:2017tud,Caputo:2021rux,Hoof:2022xbe,Diamond:2023scc}; 
gravitational waves~\cite{Diamond:2023cto}; the freeze-in ALP with mass $\lesssim 100$ MeV~\cite{Langhoff:2022bij} (when the present $a$-ALP is supposed to be produced by the freeze-in mechanism with 
the vanishing initial abundance). 
See also the public website at {\url{ https://cajohare.github.io/AxionLimits}} or {\url{ https://cajohare.github.io/AxionLimits/docs/ap.html}}.  
The current most stringent upper bounds on $g_{a\gamma\gamma}$ at at around the ALP mass on the QCD scale are read off as 
\begin{align}
    |g_{a\gamma\gamma}| & \lesssim 10^{-3} (10^{-4})\,{\rm GeV}^{-1} \,, 
    \qquad {\rm for} \qquad {0.14  \,{\rm GeV} \, (= m_\pi) < m_a < 5 (3)\, {\rm GeV}} 
    \, \text{\cite{Belle-II:2020jti}} (\text{\cite{BESIII:2022rzz,BESIII:2024hdv}}) 
    \,, \notag\\ 
    |g_{a\gamma\gamma}| & \lesssim 10^{-4}\,{\rm GeV}^{-1} \,, 
    \qquad {\rm for} \qquad {5 \,{\rm GeV} < m_a < 20\, {\rm GeV}} 
    \, \text{\cite{CMS:2018erd}}
    \,. 
\end{align}
From Eq.(\ref{Cgammagamma}), these limits can be interpreted 
in terms of $f_a$ as 
\begin{align}
     f_a & \gtrsim  2.3 (0.23) \,{\rm GeV} \times \left( \frac{|C_{\gamma\gamma}^{\rm eff}|}{1} \right) \,, 
    \qquad {\rm for} \qquad {0.14  \,{\rm GeV} \, (= m_\pi) < m_a < 5 (3)\, {\rm GeV}} 
    \, \text{\cite{Belle-II:2020jti}}(\text{\cite{BESIII:2022rzz,BESIII:2024hdv}})  
    \,, \notag\\ 
     f_a & \gtrsim  23 \,{\rm GeV} \times \left( \frac{|C_{\gamma\gamma}^{\rm eff}|}{1} \right)
    \qquad {\rm for} \qquad {5 \,{\rm GeV} < m_a < 20\, {\rm GeV}} 
    \, \text{\cite{CMS:2018erd}}
    \,. 
\end{align}

Above the mass $> 20$ GeV (up $\sim$ 100 GeV), the limit from~\cite{CMS:2018erd} gets milder  
to allow the ALP with $|g_{a\gamma\gamma}| \gtrsim 10^{-4} \, {\rm GeV}^{-1}$, instead, however, 
the OPAL in the LEP-II experiment~\cite{OPAL:2002vhf} and the LHC- ATLAS limits \cite{ATLAS:2015rsn} still exclude the case with $|g_{a\gamma\gamma}| \gtrsim 10^{-3} \, {\rm GeV}^{-1}$. 
Thus, the ALP in this mass range with $f_a/|C_{\gamma\gamma}^{\rm eff}|$ of the QCD scale 
has been severely ruled out.

For the ALP with mass below the QCD pion mass $\sim 140$ MeV, 
the electron beam damp experiments give the stringent bound on $|g_{a\gamma\gamma}|$. 
Up to the lowest $a$-ALP mass set by the QCD axion mass scale $\sim 30$ MeV in 
the limit of $m_\chi =0$ (See Eq.(\ref{ma:ref})), we find  
$10^{-5}\, {\rm GeV} < |g_{a\gamma\gamma}| < 10^{-3}\,{\rm GeV}$ 
for $30\, {\rm MeV} < m_a < 140\, {\rm MeV}$ 
(See, e.g., the summary plot in Fig.5 of \cite{Belle-II:2020jti}).

Since $f_a \sim f_{\pi_d}$ as in Eq.(\ref{fa}), which is supposed to be $\sim 100-200$ MeV, 
the upper bounds imply that 
the ALP needs to have $|C_{\gamma\gamma}^{\rm eff}| \lesssim 10^{-1}$ 
for the lower mass, and $|C_{\gamma\gamma}^{\rm eff}| \lesssim 10^{-2}$ for the 
higher mass. 
Look at Eq.(\ref{ma:ref}), which implies $m_a \gtrsim 30 -60$ MeV for 
$f_a = 100-200$ MeV to be around the scale of $f_\pi$ in QCD. 
Therefore, 
as long as the ALP is allowed to couple to the 
lightest two quarks $u$ and $d$, 
$|C_{\gamma\gamma}^{\rm eff}|$ is necessarily 
dominated by their nondecoupling loop contributions,  
which yield $|C_{\gamma\gamma}^{\rm eff}| = {\cal O}(1)$. 
Thus, the present $a$-ALP can survive the current experimental bound 
only when  
\begin{itemize}
    \item[i)]
    $m_a \sim m_\pi \sim 140$ MeV, irrespective to allowing the coupling to the lighter quarks or not; this flavorful pionic $a$-ALP with mass $\sim 140$ MeV can be probed 
in the flavor-violating processes in $B$ physics in a way similar to 
the one addressed in the literature~\cite{Ishida:2020oxl}. 
Allowing the CP violation in the CKM and the $U_R^d$ rotations, 
the ALP can be hunted by the time-dependent CP asymmetry in 
$B \to K \pi \gamma$;  

    \item[ii)] 
    $m_a \neq 140$ MeV, assuming the ALP predominantly coupled to $t$ and/or $b$ quarks, which highly suppresses the loop function in $|C_{\gamma\gamma}^{\rm eff}|$, hence $|g_{a\gamma\gamma}|$ can be as small as the current upper bounds.  
\end{itemize}

The case ii) needs more clarification, that we shall discuss in more details below. 
The $a$-ALP diagonal couplings to the lighter quarks relevant to 
the $a$-photon coupling $C_{\gamma\gamma}^{\rm eff}$ are 
given as $(g_A^u)_{ii}$ and $(g_A^d)_{ii}$ for $i=1,2$ in Eq.(\ref{gVA}). 
Using the CKM matrix elements, $V_{\rm CKM}$ (in magnitude), available from the experimental data~\cite{ParticleDataGroup:2024cfk}:   
\begin{align}
V_{\rm CKM}= 
    \left( 
    \begin{array}{ccc} 
0.974 & 0.225 & 3.8\times 10^{-3} \\ 
-0.221 & 0.987 & 4.1\times 10^{-2}  \\ 
8\times 10^{-3} & -3.88\times 10^{-2} & 1 
    \end{array}
    \right) 
  \sim
  \left( 
    \begin{array}{ccc} 
1 & \lambda & \lambda^3 \\ 
-\lambda & 1 & \lambda^2  \\ 
\lambda^3 & -\lambda^2 & 1 
    \end{array}
    \right) 
    \,, 
\end{align}
with the Cabibbo angle $\lambda \sim 0.2$, and quark mass values (MS bar masses), we roughly evaluate $(g_A^u)_{ii}$ and $(g_A^d)_{ii}$ as 
\begin{align}
    \frac{(g_A^u)_{11}}{m_u} &= c_{12}^u c_{13}^u \,, \notag\\ 
    \frac{(g_A^u)_{22}}{m_c} &= c_{12}^u c_{23}^u - s_{12}^u s_{13}^u s_{23}^u  \,, \notag\\  
    \frac{(g_A^u)_{33}}{m_t} &= c_{13}^u c_{23}^u \,, \notag\\
    \frac{(g_A^d)_{11}}{m_d} &= 0.974 c_{12}^d c_{13}^d - 4.47 \left(-s_{12}^d c_{23}^d - c_{12}^d s_{13}^d s_{23}^d\right) + 3.38\left(s_{12}^d s_{23}^d - c_{12}^d s_{13}^d c_{23}^d\right) 
    \,, \notag\\ 
    \frac{(g_A^d)_{22}}{m_s} &= 0.011 s_{12}^d c_{13}^d +0.987 \left(c_{12}^d c_{23}^d - s_{12}^d s_{13}^d s_{23}^d\right) - 1.87\left(-c_{12}^d s_{23}^d - s_{12}^d s_{13}^d c_{23}^d\right)  
    \,, \notag\\
    \frac{(g_A^d)_{33}}{m_b} &= 9\times 10^{-6} s_{13}^d  +0.087 c_{13}^d s_{23}^d - c_{13}^d c_{23}^d  
    \,.\label{Yukawa-a}
\end{align}
Taking $c_{12}^u=s_{13}^u=0$, we have $(g_A^u)_{11,22}=0$, i.e., no loop contributions to $C_{\gamma\gamma}^{\rm eff}$ from $u$ and $c$ quarks. 
However, it turns out that the $d$ and/or $s$ quark contributions are necessarily 
left no matter what mixing angle choice is applied, due to nonzero Cabibbo angle $\lambda$. 
Hence the case ii) is excluded.

We thus conclude that the case i) with only the pionic $a$-ALP is phenomenologically 
allowed. 
Note that $m_\chi$ in Eq.(\ref{ma:ref}) is necessary to be present, otherwise the $a$-ALP mass cannot be lifted up to~$~\sim 140$ MeV. 
The pionic ALP can be realized, e.g., when 
\begin{align}
    m_a  &\simeq 140 \,{\rm MeV} \times 
    \left( \frac{\langle - \bar{F}F \rangle}{(600\,{\rm MeV)^3}} \right)^{1/2} 
    \cdot \left( \frac{200\,{\rm MeV}}{f_a} \right)\cdot 
    \left( \frac{m_\chi}{0.45\,{\rm MeV}} \right)^{1/2}  
\,. \label{ma:pion}
\end{align}
Note also that the $m_\chi$ term does not directly give the mass for the dark quarks, 
hence keeps the approximate realization of the dark chiral symmetry without spoiling 
the standard description of the chiral dynamics, that allows the chiral perturbation to still work also in dQCD as in QCD.


The $a$-pionic ALP is required to have the same particle identifier as 
the QCD neutral pion, $\pi^0$, when shows up in the detector. 
This means the $a$-pionic ALP needs to have the same lifetime as $\pi^0$s, 
and gives the constraint conditions on the couplings in Eq.(\ref{Yukawa-a}). 
Moreover, the coupling between $a$ and $u,d$ quarks should be zero, 
i.e., $\frac{(g_A^u)_{11}}{m_u}=\frac{(g_A^u)_{11}}{m_d} \to 0$, 
otherwise the $a$ can mix with $\pi^0$ by the ideal mixing (with the angle of $\pi/4$), which will be ruled out by the current observations of $\pi^0$ physics.  
Thus three conditions in total are imposed in Eq.(\ref{Yukawa-a}), which 
leaves other three angle parameters. 
The flavor physics of this third-generation-philic ALP would be worth exploring in more details, in a way similar to the analysis done in~\cite{Ishida:2020oxl,Cui:2021dkr}, to be pursued elsewhere.

\subsection{The ALP: $P$} 

Putting the $a$-mass formula in Eqs.(\ref{ma:ref}) with 
the pionic ALP identifier (\ref{ma:pion}) into the $P$-mass formula in Eq.(\ref{ma:ref}), we find   
\begin{align}
    m^2_P 
   &\simeq \frac{1}{4 v_S^2} (- \chi_{\rm top}^{\rm QCD}) 
   \notag\\
    & \simeq \left(0.1\,{\rm eV} \times \left( \frac{|\chi_{\rm top}^{\rm QCD}|}{(76\,{\rm MeV})^4} \right)^{\frac{1}{2}}
     \cdot\left( \frac{2.8\times{10^5}\,{\rm TeV}}{v_S} \right)\right)^2
   \,. \label{P:mass}
\end{align}
This implies that the $P$-ALP phenomenologically plays 
the role identical to the typical QCD-invisible axion, with 
the intrinsic PQ scale $(2v_S)$.

The $P$-ALP can decay to diphoton via mixing with 
the $a$-ALP and/or $\eta'$ after passing through mixing with $\eta_d$ 
(see also Eq.(\ref{V})). 
Since we are interested in a sort of an ultralight $P$-ALP, as in Eq.(\ref{P:mass}), 
the $a$-$P$ mixing induced $P-\gamma-\gamma$ coupling will be 
gigantically suppressed. This is due to the decoupling feature of QCD quark 
loop contributions in $C_{\gamma\gamma}^{\rm eff}$, Eq.(\ref{Cgammagamma}): 
$B_1(x) \to 0$ as $x = 4m_q^2/m_P^2 \to \infty$. More precisely, $C_{\gamma\gamma}^{\rm eff}$ for the $P$-ALP goes dumped scaling with $(m_P)^3$. 
This also reflects the fact that the $a$-ALP or $P$-ALP is not (mainly) composite of QCD quarks. 
On the other hand, $\eta'-\gamma - \gamma$ coupling does not follow 
such a decoupling, because this is associated with 
the non-Abelian Wess-Zumino-Witten (WZW) anomaly~\cite{Wess:1971yu,Witten:1983tx}, or equivalently 
due to the anomaly matching in QCD. 
Thus the $P$-ALP coupling to diphoton dominantly arises from 
the mixing with $\eta'$ via mixing with $\eta_d$, in Eq.(\ref{V}).

The $P$-ALP coupling to diphoton is then evaluated as 
\begin{align}
    g_{P\gamma\gamma}& \approx g_{P\gamma\gamma}\Bigg|_{\eta' - \eta_d}
    =m_P \cdot \frac{\sqrt{6}m_a^2 \sqrt{|\chi_{\rm top}^{\rm QCD}|}}{2 m_{\eta'}^2 f_\pi m_{\eta_d}^2 } 
    \cdot g_{\eta'\gamma\gamma}(p^2\simeq 0) 
    \notag\\
    & \simeq 4 \times 10^{-10} \times \left(\frac{m_P}{1\,\rm{keV}}\right)\cdot \left(\frac{2\,\rm{GeV}}{m_{\eta_d}}\right)^2 
    \cdot g_{\eta'\gamma\gamma}(p^2\simeq 0)  
    \,,  \label{gPgg}
\end{align}
where the nondecoupling $\eta'-\gamma-\gamma$ coupling at zero momentum transfer (with $m_P$ approximated to be zero) can be read off from the WZW term~\cite{Wess:1971yu,Witten:1983tx}, 
\begin{align}
    \frac{1}{4}g_{\eta'\gamma\gamma}\eta'\tilde{F}\cdot F\,,
\end{align}
with 
\begin{align}
    g_{\eta'\gamma\gamma}=\frac{2\sqrt{2}}{\sqrt3} \frac{\alpha_{\rm em}}{\pi f_{\eta'}}\,. 
\end{align}
From the Particle Data Group~\cite{ParticleDataGroup:2022pth}, we have the partial decay rate $\Gamma[\eta' \to \gamma\gamma] \simeq 4.3 \, \rm{keV}$, for the mass $m_{\eta'}\simeq 958\, \rm{MeV}$. 
Then we get $f_{\eta'}\simeq 120\, \rm{MeV}$ and $ g_{\eta'\gamma\gamma} \simeq 
3.16 \times 10^{-2} \, \rm{GeV}^{-1}$. 
Thus the $P$-ALP coupling to diphoton is evaluated as 
\begin{align}
    g_{P\gamma\gamma}\simeq 10^{-11}\,{\rm{GeV}^{-1}} \times \left(\frac{m_P}{1\,\rm{keV}}\right)\cdot \left(\frac{2\,\rm{GeV}}{m_{\eta_d}}\right)^2 \,.
\label{Pgg}
\end{align}

We also estimate the $P \to \gamma\gamma$ decay rate (which corresponds to the inverse of the lifetime),   
\begin{align}
    \Gamma_P 
    &\simeq 10^{-33} \,{\rm eV}\times \left(\frac{m_P}{1\, {\rm keV}}\right)^5 \cdot \left(\frac{2\rm{GeV}}{m_{\eta_d}}\right)^4 
\,. 
    \label{lifetime}
\end{align}
The lifetime $\tau_P = 1/\Gamma[P\to \gamma\gamma] \sim 10^{57}$s  
is much longer than the age of the universe $\sim 10^{17}$s, hence this $P$-ALP 
is a dark matter. The relic abundance can be produced via the coherent oscillation just like the QCD axion dark matter. We will come back to the issue on the dark matter production later in 
the section of Summary and discussions.

The current ALP-photon coupling limits 
are available at {\url{ https://cajohare.github.io/AxionLimits}}, which 
gives the constraint on the $m_P-g_{P\gamma\gamma}$ trajectory in Eq.(\ref{Pgg}) 
as $m_P \lesssim 1$ eV. 
The signal trajectory will be however far off  
the future prospected observation regimes of hellioscopes. 
A recent indirect detection proposal via the Subaru telescope~\cite{Yin:2023uwf} will not be capable to probe 
this ultralight $P$-ALP.  
This result is tied with the size of $m_a$: if $m_a$ were larger, some hellioscopes 
(e.g., the prospected LAMPOST (Light $A'$ Multilayer Periodic Optical SNSPD Target) hellioscope observation) could probe 
the $P$-ALP around $m_P \sim 0.1$ eV.  
Thus the $P$-ALP is constrained to be invisible in order to predict the pionic $a$-ALP.

\subsection{The ALP: $\eta_d$}

The $\eta_d$-ALP strongly couples to other dQCD hadrons. 
The $\eta_d$-hadronic couplings are parametrized in terms of the linear-sigma model 
description. Let the sigma-model field be $M_d$, 
and parametrize it as $M_d = {\cal S}+i {\cal P} $ with the scalar-meson matrix field ${\cal S}$ and the pseudoscalar-meson matrix field ${\cal P}$, which take a $4 \times 4$ matrix form. 
This $M_d$ acts as the interpolating field 
for the dQCD fermion bilinear: $M_d^{ij} \approx \bar{F}_{R}^j F_{L}^i$, hence 
reflects the chiral $U(4)_L \times U(4)_R$ symmetry, just like $U_d$ in Eq.(\ref{Ud}). 
The nonzero VEV of ${\cal S}$, aligned to be vectorial ($\propto f_{\pi_d}$), realizes 
the desired spontaneous breaking, $U(4)_L \times U(4)_R \to U(4)_V$.

In the linear-sigma model description, 
the $U(4)_L \times U(4)_R$ symmetric potential for $M$ takes the form 
\begin{align} 
 V(M_d) =\lambda_1 {\rm tr}[(M_d^\dag M_d)^2]  + \lambda_2 ({\rm tr}[M_d^\dag M_d])^2 
 \,, \label{LSM:etad}
\end{align}
where $\eta_d$ is embedded in the singlet representation part of ${\cal P}$. 
Allowing the VEV of ${\cal S}$ to be $\propto f_{\pi_d} \cdot 1_{4 \times 4}$, 
the trilinear coupling with $\eta_d$ can be generated from there, which 
would be in part relevant to clarifying the main decay channel of $\eta_d$. 
Note that due to the strong QCD gauge interaction, 
the residual $U(4)_V$ symmetry is explicitly broken down to be split into the 
$U(3)_V \times U(1)_V$ structure, where the $SU(3)$ portion in $U(3)_V$ is identified as 
the $SU(3)_c$ of QCD. 
Hence the QCD-colored dQCD hadrons 
eventually get massive enough to be on the same order as 
$\eta_d$ mass. 
Thus only the QCD-color singlet mesons, the $a$-ALP and its scalar meson counterpart ${\cal S}_a$, would be crucial in Eq.(\ref{LSM:etad}) to investigate the 
$\eta_d$ couplings to lighter dQCD hadrons. 
We then find the relevant trilinear coupling term, 
\begin{align} 
{\cal L}_{\eta_d {\cal S}_a a} \sim \frac{\Lambda_d^2}{f_{\pi_d}} \cdot \eta_d \, a \, {\cal S}_a 
 \,, 
\end{align}
where we have roughly evaluated $\lambda_{1,2} \sim (\Lambda_d/f_{\pi_d})^2$. 
The ${\cal S}_a$ mass can be smaller than the $\eta_d$ mass to be 
as small as 140 MeV of the $a$-ALP, or higher. 
Thereby, the $\eta_d$-ALP can mainly decay to ${\cal S}_a$ and the $a$-ALP 
with the strong coupling length $\sim (4\pi)^2 f_{\pi_d}$, like 
the QCD sigma meson decay to pion pairs. 
The $\eta_d \to {\cal S}_a + a$ decay rate can be evaluated as 
\begin{align}
    \Gamma[\eta_d \to S_a + a] \sim \frac{m_{\eta_d}^3}{16 \pi f_{\pi_d}^2} 
    \left( 1 - \frac{m_{{\cal S}_a}^2}{m_{\eta_d}^2} \right)
\,, \label{main:decay}
\end{align}
where we have roughly replaced $\Lambda_d$ with the $\eta_d$ mass $m_{\eta_d}$ and taken the $a$-ALP mass to be zero.

The $\eta_d$-ALP can also couple to diphoton, which 
is generated via mixing with $\eta'$ in the potential, Eq.(\ref{V}): 
\begin{align}
    g_{\eta_d\gamma\gamma}&=\frac{3\sqrt{3} |\chi_{\rm{top}}^{\rm QCD}|}{4f_{\pi} f_{\pi_d}} \cdot \frac{1}{m_{\eta_d}^2-m_{\eta'}^2}
    \cdot g_{\eta'\gamma\gamma}
    \notag\\
    & \simeq  { 7.2\times 10^{-5} \, \rm{GeV}^{-1}} \times \left(\frac{200\,\rm{MeV}}{f_{\pi_d}}\right) \cdot \left(\frac{1 \,\rm{GeV} ^2 }{m_{\eta_d}^2-m_{\eta'}^2}\right)\,. 
\label{etad-g-g}
\end{align} 
Here the mixing angle $\theta_{\eta'\mathchar`-\eta_d}$ between $\eta_d$ and $\eta'$ reads off from Eq.(\ref{etad-g-g}) to be $\sim 10^{-2}$ at $f_{\pi_d}=100$ MeV and $m_{\eta_d}=1$ GeV, which is to be smaller as $f_{\pi_d}$ or $m_{\eta_d}$ gets larger. In comparison, the current accuracy of measuring the $\eta'$ coupling is about 4.5\% (read off from the accuracy of the full width)~\cite{ParticleDataGroup:2022pth}, which 
is greater than the size of the maximal possible deviation reached when $f_{\pi_d}=100$ MeV and $m_{\eta_d}=1$ GeV: $(1 - \Gamma_{\eta'}/\Gamma_{\eta'}^{\exp}) = 1 - |\cos \theta_{\eta'\mathchar`-\eta_d}|^2 \sim 10^{-3}$. Thus the present $\eta_d$-ALP for $f_{\pi_d} \gtrsim 100$ MeV 
currently survives not to spoil the $\eta'$ physics.

From Eq.(\ref{etad-g-g}) 
the $\eta_d \to 2 \gamma$ decay rate is evaluated as 
\begin{align} 
 \Gamma[\eta_d \to 2 \gamma] = \frac{g_{\eta_d \gamma\gamma}^2 m_{\eta_d}^3}{64\pi}
\,,  
\end{align} 
which is, when combined with Eq.(\ref{main:decay}), 
read as the branching fraction, 
\begin{align}
    {\rm Br}[\eta_d \to 2 \gamma] 
    \sim 10^{-11} - 10^{-10} 
    \,, 
\end{align}
for $m_{\eta_d} \sim 1 - 2 $ GeV, $f_{\pi_d} \sim 200$ MeV, and $m_{{\cal S}_a} < m_{\eta_d}$. 
This branching fraction is too tiny to be sensitive to 
the ongoing Belle II and BESIII experiments, where $e^- e^+ \to 3 \gamma$ events have been 
measured and the upper limit on the ALP-photon coupling assuming ${\rm Br}[{\rm ALP} \to 2 \gamma] =1$. 
Although $\eta_d$ is thus unlikely to be seen via the direct diphoton channel, 
it can have multi-photon decay channel via the main decay process, in Eq.(\ref{main:decay}):   
$\eta_d \to {\cal S}_a + a$ with $S_a$ being a missing energy (or decaying $\to 2 \gamma)$ and 
$a \to 2 \gamma$. Here 
${\cal S}_a$ would decay to $\gamma\gamma$ similarly to the $a$-ALP, 
but the coupling to diphoton would be highly suppressed because it arises at 
least three-loop level (passing through $S_a \to gg$ via the QCD-charged dQCD quark loops, 
$gg \to \gamma\gamma$ via the ordinary QCD quark loops). 
This signal may be contaminated with the typical $3\gamma$ events at Belle II, BESIII,  
and EIC experiments. Thus, at any rate, the $\eta_d$-ALP can have the sensitivity to 
be probed at those collider experiments.

At the Belle II, 
the currently available observation limit around the target mass and photon coupling 
comes from $496\,{\rm pb}^{-1}$ data. 
This experiment has placed the upper bound on the photon coupling,  
$g_{a \gamma \gamma} \lesssim 10^{-3}\,{\rm GeV}^{-1}$ around the mass 
in a range of $1 -5$ GeV~\cite{Belle-II:2020jti}. 
The Belle II future prospects with $20\,{\rm fb}^{-1}$ and $50\,{\rm ab}^{-1}$ data 
will reach $g_{a \gamma \gamma} \sim 10^{-4}\,{\rm GeV}^{-1}$ at the mass around $1- 5$ GeV~\cite{Dolan:2017osp}, 
which can probe $\eta_d$ via the $3 \gamma$ signals. 
The existing limits around the target mass range are available from the literature~\cite{Bauer:2017ris}.

Recently the BESIII experiments  
have reported the updated bounds on the ALP photon coupling 
based on the $J/\psi$ events~\cite{BESIII:2024hdv}, which has overall improved 
the previous bounds derived from the $\psi(3686)$ events~\cite{BESIII:2022rzz}. 
The ALP photon coupling has thus been constrained to be $g_{a \gamma \gamma} \lesssim 3 \times 10^{-4}\,{\rm GeV}^{-1}$ at around the mass $\sim 1$ GeV.

The EIC experiment projections~\cite{Balkin:2023gya} can also be sensitive to  
the present $\eta_d$-ALP decaying photons. 
The upcoming $10\,{\rm fb}^{-1}$ data on the promptly decaying ALP with mass 
around $1 - 10$ GeV can probe the photon coupling up to $\gtrsim (3 - 7) \times 10^{-5} \, {\rm GeV}^{-1}$. 
When the ALP is long-lived, the displaced vertex search at the EIC 
can have a higher sensitivity for the ALP with mass 
around 1 GeV, which expects to probe the photon coupling up to $\gtrsim 10^{-6} \, {\rm GeV}^{-1}$ 
at the luminosity $\sim 10 - 100 \, {\rm fb}^{-1}$.

\begin{figure}[t]
  \begin{center}
   \includegraphics[width=10cm]{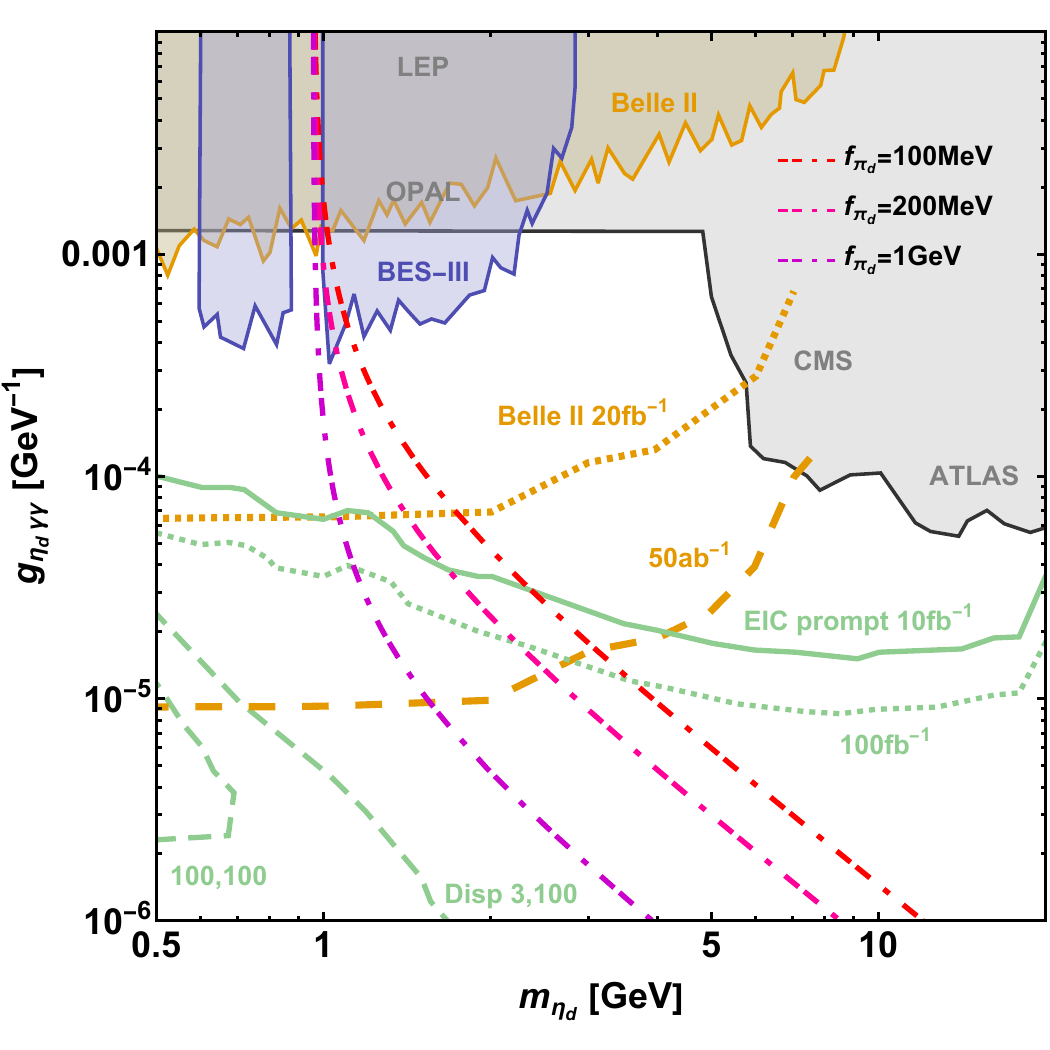}
  \end{center}   
\caption{
The current limits on the $\eta_d$-ALP mass and photon coupling, quoted from 
the  public website {\url{ https://cajohare.github.io/AxionLimits/docs/ap.html}}. 
The predicted $\eta_d$-ALP trajectory has been drawn varying $f_{\pi_d}$ in 
a range of 100 MeV$-$1 GeV. 
The sharp blowing-up trend at around the QCD $\eta'$ mass scale 
reflects the resonant enhancement structure in the mixing induced coupling form 
as in Eq.(\ref{etad-g-g}) (without finite width contributions). 
Also have been shown the future prospected 
detection sensitivity curves from the Belle II experiments with 
$20\,{\rm fb}^{-1}$ (dotted curve) and $50\,{\rm fb}^{-1}$ (dashed curve)~\cite{Belle-II:2020jti} and 
the projection curves in the EIC experiment at $10\,{\rm fb}^{-1}$ (solid curve) and $100
\,{\rm fb}^{-1}$ (dotted)~\cite{Balkin:2023gya}, in which 
the prompt ALP decaying to diphoton (denoted as ``prompt" in the figure) 
and long-lived ALP searches using the displaced vertex (marked as ``Disp") are included . 
The ``Disp 3, 100" and ``Disp 100, 100" (dashed curves) respectively stand for 
the cases with the expected number of signal events being 3 and 100 at the integrated 
luminosity of $100\,{\rm fb}^{-1}$~\cite{Balkin:2023gya}.  
}  
\label{Figg6yy}
\end{figure}

Figure~\ref{Figg6yy} shows the exclusion plots on the mass and the photon coupling space  
together with the prospected projection curves in light of the Belle II and EIC experiments 
including the detection via the displaced vertex. 
The $\eta_d$-ALP with the mass in a range $1.2\, \rm{GeV} - 3.2 \, \rm{GeV}$ can have high sensitivities to be probed by the future Belle II and EIC (in the category of the prompt ALP decaying to diphoton).

\section{Summary and discussions}

In summary, 
by addressing the strong CP problem,  we have discussed an extension of a phenomenologically and cosmologically viable QCD $\times$ dQCD scenario in a view of the QCD-scale baryogenesis~\cite{Wang:2022ygk,Wang:2024tda}. 
A couple of predicted (partially) composite ALPs successfully settle the QCD and dQCD vacua to be free from CP phases by the hybrid relaxation.  
We have found that this extension improves 
the predictability of the scenario of this class. 
The QCD-scale composite axion ($a$) is phenomenologically required to mimic the QCD pion (with the same mass $\sim 140$ MeV and the same lifetime), but can generically be flavorful due to the intrinsically minimal flavor violation, which is potentially testable via the induced flavor changing processes at experiments.   
The Belle II and EIC (and also potentially BESIII) experiments can have a high enough 
sensitivity to probe the dark $\eta'$-like meson ($\eta_d$) in the diphoton channel, which arises from the mixing dominantly with the QCD $\eta'$ consistently with the successful QCD hadron physics.

In addition to the $P$-ALP, predicted as a partner for the axion relaxation, 
the present dQCD generates dark baryons as another dark matter candidate. 
As in the case of the ordinary baryons, 
this $n_d$ could annihilate into the lightest meson pairs, i.e., via $n_d \bar{n}_d \to \pi_d \pi_d$, 
or more multiple $\eta_d$ states (and also into the QCD $\sigma$-like $(\sigma_d)$ states), 
which would determine the freeze out of the number density of $n_d$ in the thermal history.

A rough estimation of the thermal relic abundance of $n_d$ 
could be done by assuming the standard freeze-out scenario for this annihilation 
with the size of the classical cross section $\langle \sigma v \rangle \sim 4 \pi/m_{n_d}^2$. 
This crude approximation could be justified in the  
large-$N_c$ and -$N_d$ limit, where the dark baryon behaves like almost 
static, nonrelativistic, and a classical rigid body with finite 
radius (i.e. impact parameter) of ${\cal O}(1/m_{n_d})$. 
The thermal relic abundance is then evaluated as 
$\Omega_{n_d} h^2 \sim \frac{10^9 x_{\rm FO}}{\sqrt{g_*(T_{\rm FO})} M_{\rm p} {\rm GeV} \langle \sigma v \rangle}$, where $x_{\rm FO} = m_{n_d}/T_{\rm FO}$ with $T_{\rm FO}$ being the 
freeze-out temperature; $M_{\rm pl}$ is the Planck scale $\sim 10^{18}$ GeV; 
$g_*(T_{\rm FO})$ the effective degree of freedom of relativistic particle at $T=T_{\rm FO}$. 
The standard freeze-out scenario gives $x_{\rm FO}\sim 20$, and $g_*(T_{\rm FO}) = {\cal O}(50)$ at $T$ below $1$ GeV~\cite{ParticleDataGroup:2022pth}. 
The relic abundance of $n_d$ is then estimated to be of ${\cal O}(10^{-9})$ for 
$m_{n_d} \sim 2$ GeV, which 
is compared with the observed abundance of the cold dark matter today $\sim 0.1$.

Still, however, the nonperturbative and nonthermal production is possible to 
trigger by the nonadiabatic preheating of $\sigma_d$, which may be realized 
due to the (little) supercooling and tunneling from the false and true vacua. 
This type of the preheating could also generate the gravitational waves~\cite{Dufaux:2010cf,Adshead:2019lbr,Easther:2006gt,Garcia-Bellido:2007nns,Garcia-Bellido:2007fiu,Dufaux:2007pt,Dufaux:2008dn,Bethke:2013aba,Bethke:2013vca,Figueroa:2017vfa,Adshead:2018doq,Adshead:2019igv,Cui:2021are,Zhang:2024ggn,Cosme:2022htl}, which might be more involved due to another source provided from the first-order dark chiral phase transition. 
Moreover, the primordial black holes could also be generated by the dark chiral phase transition, to be another dark matter today~\cite{Hawking:1982ga,Lewicki:2019gmv,Jung:2021mku,Kodama:1982sf,Hall:1989hr,Kusenko:2020pcg,Liu:2021svg,Hashino:2021qoq,Hashino:2022tcs,He:2022amv,Kawana:2022olo,Lewicki:2023ioy,Gouttenoire:2023naa,Gouttenoire:2023pxh,Salvio:2023blb,Conaci:2024tlc,Banerjee:2023brn,Banerjee:2023vst,Banerjee:2023qya,Baldes:2023rqv,Lewicki:2024ghw,Balaji:2024rvo,Jinno:2023vnr,Flores:2024lng,Lee:1986tr,Gross:2021qgx,Baker:2021nyl,Baker:2021sno,Kawana:2021tde,Huang:2022him,Kawana:2022lba,Lu:2022jnp,Marfatia:2021hcp,Marfatia:2022jiz,Tseng:2022jta,Xie:2023cwi,Acuna:2023bkm,Lewicki:2023mik,Chen:2023oew,Kim:2023ixo,Gehrman:2023qjn,Borah:2024lml,Mazumdar:2018dfl,Kanemura:2024pae,Goncalves:2024vkj,Ai:2024cka}. 
Thus, determination of the main dark matter component today is subject to those cosmological questions, hence the detection sensitivity at the dark matter direct detection experiments are also too premature to argue. 
Those issues deserve to another publication.

The dynamical origin of the spontaneous breaking of the PQ symmetry for the $P$-ALP (that we shall hereafter call the $P$-ALP PQ symmetry) can be provided from 
the Higgs-portal dynamical scalegenesis based on the Gildener-Weinberg mechanism~\cite{Gildener:1976ih}, as in the literature~\cite{DelleRose:2019pgi}. 
This type of the scenario can realize 
the electroweak symmetry breaking and the scale generation at the same time as the $P$-ALP PQ symmetry breaking. As in the reference, the additional gravitational waves can also be 
generated via the supercooled PQ transition.  
Thus the details of this $P$-ALP PQ breaking mechanism involves 
more cosmological and phenomenological discussions. 
Combining the dQCD-induced gravitational waves with those produced from the 
$P$-ALP PQ transition (and also the dQCD preheating-induced ones as noted above), 
the predicted gravitational wave spectra could be richer.  
This issue is noteworthy to address elsewhere.

\section*{Acknowledgments} 

We are grateful to Hexu Zhang for fruitful discussions. 
This work was supported in part by the National Science Foundation of China (NSFC) under Grant No.11747308, 11975108, 12047569, 
the Seeds Funding of Jilin University (S.M.), 
and JSPS KAKENHI Grant Number 24K07023 (H.I.).

\end{document}